\documentclass[prd,nofootinbib,showpacs,twocolumn]{revtex4}
\usepackage{graphicx}
\usepackage[usenames,dvipsnames]{color}
\usepackage{amsmath,amssymb}
\usepackage{epstopdf}
\usepackage{hyperref}

\begin{document}

\title{Reviving quark nuggets as a candidate for dark matter}

\author{Abhishek Atreya}
\email{atreya@iopb.res.in}
\affiliation{Institute of Physics, Bhubaneswar, 751005, India}
\author{Anjishnu Sarkar}
\email{anjishnusarkar@gmail.com}
\affiliation{LNM Institute of Information Technology, Jaipur, Rajasthan,
302031, India}
\author{Ajit M. Srivastava}
\email{ajit@iopb.res.in}
\affiliation{Institute of Physics, Bhubaneswar, 751005, India}

\begin{abstract}
We discuss a novel mechanism for segregation of baryons and anti-baryons in 
the quark-gluon plasma phase which can lead to formation of quark and 
antiquark nuggets in the early universe, irrespective of the order of the 
quark-hadron phase transition. This happens due to
CP violating scattering of quarks and antiquarks from moving $Z(3)$
domain walls. CP violation here is spontaneous in nature and arises
from the nontrivial profile of the background gauge field ($A_{0}$ ) between
different $Z(3)$ vacua. We study the effect of this spontaneous CP 
violation on the baryon transport across the collapsing large $Z(3)$ domain 
walls (which can arise in the context of certain low energy scale inflationary 
models). Our results show that this CP violation can lead to large 
concentrations of baryons and anti-baryons in the early universe.
The quark and antiquark nuggets, formed by this alternate mechanism, 
can provide a viable dark matter candidate within standard model without 
violating any observational constraints.
\end{abstract}

\maketitle

\section{INTRODUCTION}
\label{sec:intro} 

 One of the main unsolved problems of the modern physics is the existence
of dark matter in the universe. It is usually stated that the data on 
Nucleosynthesis and CMBR does not allow baryonic dark matter.
This indeed holds true for baryons in the form of gas (e.g. hydrogen,
helium). Observational constraints from nucleosynthesis and CMBR are
very strong on such forms of baryonic matter and restrict it to
less than 20 \% of all matter/radiation in the universe (excluding
the dark energy). However, it is important to note that
these constraints do to apply if baryons are in the form of heavy 
bodies, such as quark nuggets, MACHOS, etc., provided that
such objects form before nucleosynthesis. There are separate 
strong observational constraints on MACHOS  from 
gravitational microlensing observations. In any case, it is hard to
come up with scenarios where such heavy objects could form before
nucleosynthesis. On the other hand, quark nuggets pass through all 
the observational constraints, and indeed,
these were considered promising dark matter candidates after the
pioneering work of Witten \cite{Witten:1984rs} showing the possibility of
formation of such objects in a strong first order quark-hadron transition
in the universe. There were many investigations discussing the issues
of stability of such objects 
\cite{Bhattacharjee:1993ah,Lugones:2003un,Bhattacharyya:2002vk}. It 
was generally considered
that quark nuggets (strangelets) having density above nuclear density, 
with baryon number ranging from few Thousand to $ \sim 10^{50}$ (sizes 
varying from fm to meters) can provide required dark matter. Such a 
candidate for dark matter will be extremely appealing as it does not 
require any physics beyond standard model.

  The interest in quark nuggets declined with results from lattice
gauge theory showing that a first order quark-hadron transition
is very unlikely. The transition, for the range of chemical potentials
relevant for the early universe, is most likely a crossover. Witten's
scenario of formation of quark nuggets does not work in such a case.
However, with most attempts of explaining the dark matter not meeting
any success (such as supersymmetric dark matter candidates in view
of LHC results), it is important to appreciate following points
about quark nuggets as dark matter candidates. As we mentioned
above, here one does not need any new species of particles, quarks
do the job. Secondly, any scenario of forming quark nuggets will most 
naturally fit in the QGP phase of the universe, well above radiation 
decoupling and nucleosynthesis stages. Those baryons (quarks) which form 
(heavy) quark nuggets completely decouple from the processes happening at
nucleosynthesis stage, and later on at the radiation decoupling stage.
Thus, nucleosynthesis and CMBR constraints do not apply to the
fraction of baryons in quark nuggets. Further, stability of these quark 
nuggets, especially strangelets, has been extensively discussed and it
has been argued that strangelets with baryon number of several hundred
to general quark nuggets with  baryon number of order up to $10^{50}$
may be stable up to the present stage 
\cite{Bhattacharjee:1993ah,Lugones:2003un,Bhattacharyya:2002vk}. 
The only issue then 
remains is how to form these objects when quark-hadron transition
is a cross-over. We address this issue in this paper, extending our earlier
analysis of an alternate scenario of formation of quark nuggets
without requiring any first order quark-hadron phase transition.

  We would like to emphasize that even in the absence of 
a mechanism for the formation of quark nuggets, it is important to
recognize that quark nuggets provide a viable dark matter candidate
entirely within the Standard model.
It then provides a strong motivation to search for mechanisms which
can lead to formation of such objects in the early stages of the
universe. Indeed, these exciting objects have fascinated cosmologists 
and even now there are attempts to detect these objects 
\cite{Gorham:2012hy,Astone:2013bed}.

 We briefly recall the essential physics of Witten's proposal 
\cite{Witten:1984rs} for the formation of quark nuggets. Witten proposed 
that if the universe underwent a (strong) first  order QCD phase 
transition, then localized regions of high temperature phase, trapped
between expanding hadronic bubbles, will shrink, in the process trapping 
the baryons inside them. He also argued that resulting quarks nuggets 
can be stable and survive upto the present epoch. 
In Witten's scenario, the importance of first order phase 
transition was due to the fact that it provides us with an interface between 
two region of the universe in different phases. The baryon 
transport across the phase boundary then leads to the build up of baryon 
excess in the collapsing domains. Such an interface does not exist
in a crossover or in a second order phase transition. Hence, with lattice 
QCD calculations ruling out the first order phase transition, the mechanism 
of formation of quark nuggets as proposed by Witten becomes inapplicable. 

 Our proposal for an alternate mechanism for the formation of quark nuggets
is based on this crucial ingredient of Witten's scenario, that is
the existence of an interface leading to quark/antiquark reflection.
Quark-hadron phase boundary (for a first order transition) is one
such possibility for the interface. However, in addition to this bubble
wall interface between different QCD phases, there are other possibilities 
of extended topological objects in the quark-gluon plasma (QGP) phase 
and these have been extensively discussed in the literature 
\cite{Bhattacharya:1992qb,West:1996ej,Boorstein:1994rc}. These are domain
wall defects and they arise from the spontaneous breaking of 
$Z(3)$ symmetry in the high temperature phase (QGP phase) of QCD, with
the expectation value of the Polyakov loop $L(x)$ being the order parameter
for confinement-deconfinement transition. It has been pointed 
out that there are also 
topological string defects in QGP forming at the junctions of Z(3) walls 
\cite{Layek:2005fn}. The existence of these defects can be probed in 
the ongoing relativistic heavy-ion collision experiments at BNL and at 
LHC-CERN. These are the only topological defects in a relativistic quantum 
field theory which can be probed in lab conditions with the present day 
accelerators. Detailed simulations have been performed to study the formation 
and evolution of these objects in these experiments 
\cite{Gupta:2010pp,Gupta:2011ag}. 

In the presence 
of quarks, questions have been raised on the existence of these objects 
\cite{Smilga:1993vb,Belyaev:1991cw}. However, lattice studies by Deka et al. 
\cite{Deka:2010bc} of QCD with quarks show strong possibility of the existence 
of non-trivial, metastable, $Z(3)$ vacua for high temperatures of order 
700 MeV. These high temperatures occur naturally in the early universe
and may be possible to reach at LHC. Hence, it seems plausible  that these 
defects will be naturally formed in any realistic phase transition from 
the confining phase to the QGP phase.

 The baryon inhomogeneity generation due to the reflection of 
quarks/antiquarks from Z(3) walls was first studied by some of us in the 
context of relativistic heavy-ion collision experiments (RHICE) and in
the universe, in ref \cite{Layek:2005zu}. 
For the case of the universe, it was argued in \cite{Layek:2005zu}
that these collapsing domains can concentrate enough baryon number 
(in certain late time inflationary models) to form quark nuggets thus 
providing us with an alternate scenario of quark nuggets formation in early 
universe, which is independent of the order of phase transition. 
In these works, the scattering of quarks from Z(3) walls was 
calculated by modeling the dependence of effective quark mass on the 
magnitude of the Polyakov loop order parameter $L(x)$ which did not 
distinguish between quarks and antiquarks. 

  In this paper we will extend the earlier analysis \cite{Layek:2005zu} 
by incorporating an interesting possibility arising from the 
spontaneous CP violation from Z(3) interfaces. This was first discussed
by Altes et al \cite{KorthalsAltes:1994be}, who showed that spontaneous 
CP violation can arise from  $Z(N)$ structures due to the non-trivial 
background gauge field configuration associated with the Polyakov loop.
They showed that it can lead to the localization of either quarks or 
antiquarks on the domain wall. It was also argued that it can lead to 
baryogenesis via sphaleron transition in certain extensions of the
Standard model. Same possibility of spontaneous CP violation for the 
case of QCD was also discussed in \cite{KorthalsAltes:1992us}.  
Though, in these works, the CP violating
effects were discussed primarily qualitatively, and the exact profiles
of $L(x)$ or the associated $A_0$ profiles were not calculated.

  In an earlier work \cite{Layek:2005zu} we have incorporated this spontaneous 
CP violation in the propagation of quarks and anti-quarks across the $Z(3)$ 
walls. We use the profile of Polyakov loop $L\left(\vec{x}\right)$  
between different Z(3) vacua (which was obtained by using specific 
effective potential for $L(x)$ as discussed in \cite{Pisarski:2000eq}) to 
obtain 
the profile of the background gauge field $A_0$. This background $A_0$ 
configuration acts as a potential for quarks and antiquarks causing 
non-trivial reflection of quarks from the wall. Spontaneous CP violation
arising from the background $A_0$ configuration leads to different
reflection coefficients for quarks and antiquarks. In the present 
work we study the effect of this difference in the scattering of
quarks and antiquarks from Z(3) walls  on baryon transport across the 
collapsing $Z(3)$ domain walls in the early universe. We calculate the 
transmission coefficients of quarks and antiquarks from the background 
$A_{0}$ profile and use those in the baryon transport equations. 
We show that it can lead to the generation of baryon density 
inhomogeneities, by segregating baryons and antibaryons in different 
regions of the universe near QCD phase transition epoch. (Since the background 
field is a color field, not only we get the quarks and anti-quark segregation, 
we also find that the segregation of the baryons/antibaryons depends
on the color configuration of the specific Z(3) wall.
This can have important implications in the context of the early 
universe and heavy ion experiments that could be worth pursuing.)

 Here it should be mentioned that in the present 
work we use $Z(3)$ wall profile of pure $SU(3)$ gauge theory, without dynamical 
quarks. The quark effects may not be important in the context of heavy ion 
experiments due to small length and time scales involved, but for the case of 
universe these effects will be of crucial importance. We will discuss this 
further below and argue that in case of certain inflationary models we can 
work with the domain wall profile corresponding to pure $SU(3)$ gauge theory.

  The organization of the paper is as follows. In section \ref{sec:z3}
we start by discussing the effective potential for the Polyakov 
loop and calculate the 
profile of the background gauge field $A_0$ from the profile of the order 
parameter $L\left(\vec{x}\right)$ between different Z(3) vacua 
\cite{Layek:2005fn}. In section \ref{sec:z3form} we discuss the formation of 
$Z(3)$ structures in the early universe. There we discuss in detail the 
effects of quarks in the context of inflationary cosmology and how in certain 
low energy inflationary models, these $Z(3)$ domains can survive long enough 
to have interesting cosmological implications. The formation of baryon 
inhomogeneities due to baryon transport across the $Z(3)$ walls is discussed in 
section \ref{sec:binhm}. We present our results in section \ref{sec:results}. 
Section \ref{sec:disc} presents discussions and conclusions.

\section{$Z(3)$ SYMMETRY AND SPONTANEOUS CP VIOLATION}
\label{sec:z3}

 In this section we discuss the effective potential used to study the 
confinement-deconfinement phase transition in QCD, and the basic physics 
of spontaneous CP violation from the $Z(3)$ structure. Initially we 
restrict our discussion to pure $SU(N)$ gauge theory. In pure gauge 
$SU(N)$ system, in thermal equilibrium at temperature $T$, Polyakov loop 
\cite{Polyakov:1978vu,Gross:1980br,McLerran:1981pb} is defined as 
\begin{equation} 
L(x) = \frac{1}{N}Tr\biggl[\mathbf{P} \exp\biggl(ig\int_{0}^{\beta}A_{0}
  (\vec{x},\tau)d\tau\biggr)\biggr],
\label{eq:lx}
\end{equation}
where, $\beta = T^{-1}$ and $A_{0}(\vec{x},\tau) = A_{0}^{a}(\vec{x},\tau)T^{a}, 
(a = 1,\dotsc N)$ are the $SU(N)$ gauge fields satisfying the periodic 
boundary conditions in the Euclidean time direction $\tau$, viz 
$A_{0}(\vec{x},0) = A_{0}(\vec{x},\beta)$. $T^{a}$ are the generators of 
$SU\left(N\right)$ in the fundamental representation. $\mathbf{P}$ denotes the 
path ordering in the Euclidean time $\tau$, and $g$ is the gauge coupling. 
Thermal average of the Polyakov loop, $\langle L(\vec{x})\rangle$, acts as the 
order parameter for the confinement-deconfinement phase transition. For 
brevity, we will use $l(x)$ to denote $\langle L(\vec{x})\rangle$ from now on. 
It is related to the free energy of a test quark in a pure gluonic medium, 
$l(x) \propto e^{-\beta F}$. In confined phase, the free energy of a test 
quark is infinite hence $l(x) = 0$ (i.e. system is below $T_{c}$). While 
$l(x) \neq 0$ in deconfined phase, because in the deconfined phase a test
quark has finite free energy (in other words, the system  is above the 
critical temperature $T_{c}$). Under $Z(N)$ (which is a center of $SU(N)$) 
transformation, the Polyakov Loop transforms as
\begin{equation}
L(x) \longrightarrow Z\times L(x), \qquad \text{where}~ Z = e^{i\phi},
\end{equation}
where, $\phi = 2\pi m/N$; $m = 0,1 \dotsc (N-1)$. This leads to the spontaneous 
breaking of $Z(N)$ symmetry with $N$ degenerate vacua in the deconfined phase 
or QGP phase. For QCD, $N =3$ hence it has three degenerate $Z(3)$ vacua 
resulting from the spontaneous breaking of $Z(3)$ symmetry at $T>T_{c}$. This 
leads to the formation of interfaces between regions of different $Z(N)$ vacua.
These vacua are characterized by,
\begin{equation}
l(\vec{x}) = 1, e^{i2\pi/3}, e^{i4\pi/3}.
\end{equation}

 It has been argued that these $Z(3)$ domains do not have a physical 
meaning \cite{Smilga:1993vb,Belyaev:1991cw}. As dynamical quarks do not 
respect the $Z(N)$ symmetry, their inclusion further complicates the issue. 
It has been argued that the effect of addition of quarks can be interpreted 
as the explicit breaking of $Z(N)$ symmetry, see, for example, refs.
\cite{Pisarski:2000eq,Dumitru:2000in,Dumitru:2002cf,Dumitru:2001bf}.
This leads to the lifting of degeneracy of the 
vacuum, with $l(\vec{x}) = 1$ as the true vacuum and $l(\vec{x}) = e^{i2\pi/3}, 
e^{i4\pi/3}$ as the metastable ones. We will follow this approach. This 
interpretation finds support in the recent lattice QCD studies with quarks 
\cite{Deka:2010bc}. These result strongly favor these metastable $Z(3)$ vacua 
at high temperature. These $Z(3)$ vacua can have important consequences in 
the case of early universe where these high temperatures occur quite naturally. 
However, for the time being we will consider the pure gauge case 
(i.e degenerate $Z(3)$ vacua) because our emphasis here is on the 
interesting physics due to the spontaneous CP violation in the reflection of 
quarks and antiquarks from $Z(3)$ walls which leads to the segregation of 
baryons and anti-baryons in early universe. This aspect is independent of the 
explicit symmetry breaking due to quark effects. We will discuss the effects 
of quarks again when we discuss the formation of $Z(3)$ networks in the next 
section (Section \ref{sec:z3form})

\begin{figure}
\begin{center}
\includegraphics[width=0.35\textwidth]{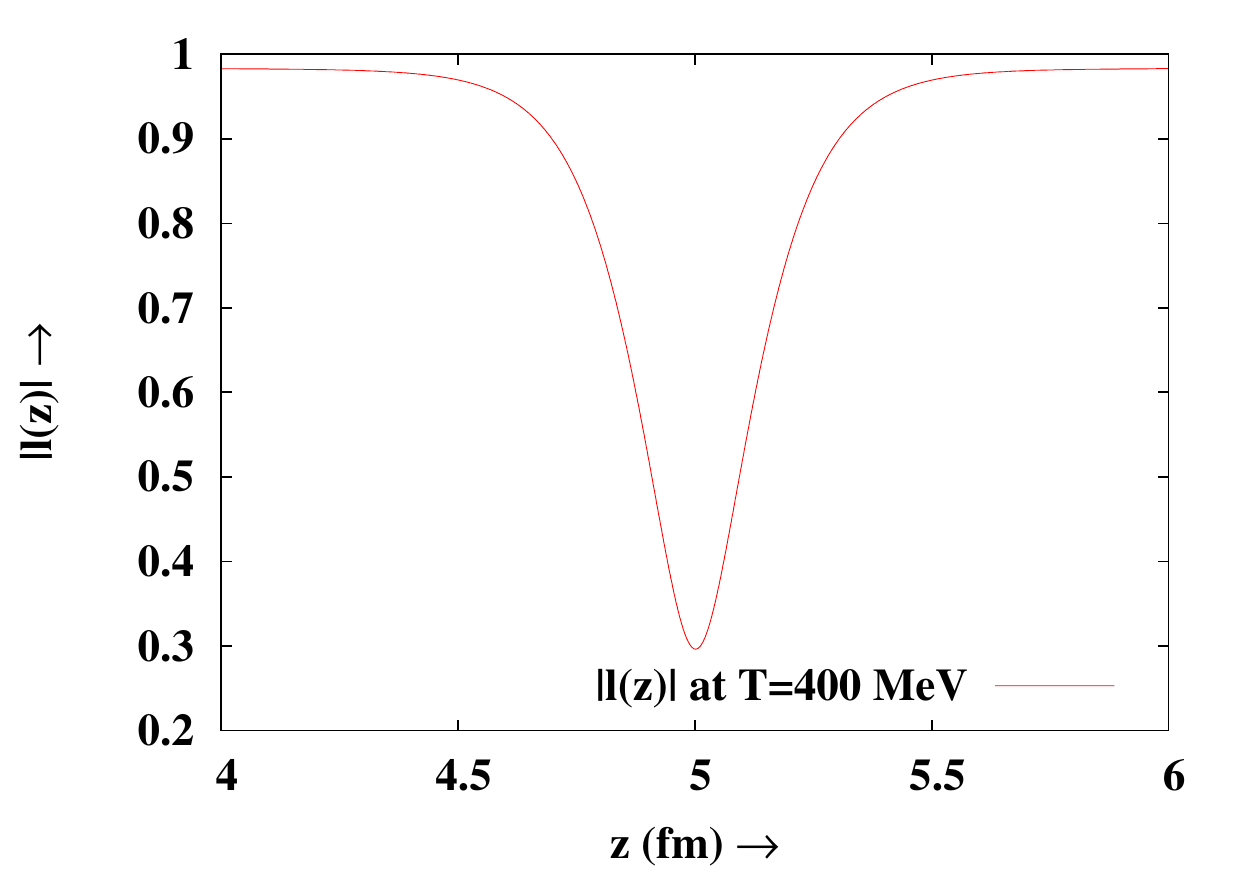}
\caption{Variation of $|l\left(x\right)|$  between different $Z(3)$ vacua 
for $T=400$ MeV. The profile is same between any two interface.}
\label{fig:lprfl}
\end{center} 
\end{figure}

 An effective potential for Polyakov loop, in the spirit of Landau-Ginzberg 
theory of phase transitions, was proposed by Pisarski \cite{Pisarski:2000eq}. 
The Lagrangian density is given as
\begin{equation} 
\mathcal{L} = \frac{N}{g^{2}}\vert \partial_{\mu}l\vert^{2}T^{2} - V(l),
\label{eq:lagrangian}
\end{equation}
 where $N=3$ for QCD. $T^{2}$ is multiplied with the first term to
give the correct dimensions to the kinetic term. $V(l)$ is the potential term 
that has the form
\begin{equation}
V(l) = \biggl(-\frac{b_2}{2}|l|^{2} - \frac{b_3}{6}\Bigl(l^{3} 
+ (l^{*})^{3}\Bigr) + \frac{1}{4}(|l|^{2})^{2}\biggr)b_4T^{4}.
\label{eq:pispot}
\end{equation}
When $T > T_{c}$ (i.e $l(x) \neq 0$), the cubic term in the above potential 
gives rise to $\cos(3\theta)$ term (by writing $l(x) = |l(x)|e^{i\theta}$), 
that leads to three degenerate $Z(3)$ vacua.  In 
ref \cite{Dumitru:2000in,Dumitru:2002cf,Dumitru:2001bf}, the coefficients 
$b_{2}$, $b_{3}$ and $b_{4}$ are fixed using lattice results for the pressure 
and energy density for pure SU(3) gauge theory 
\cite{Boyd:1996bx,Okamoto:1999hi}. $b_{2}$ is given by 
$b_2 = \left( 1-1.11/x \right) \left(1+0.265/x\right)^{2}\left(1+0.300/x
\right)^{3} - 0.478$, where $x=T/T_{c}$ with $T_{c}\sim182$ MeV. The other 
parameters are $b_3 = 2.0$ and $b_4 = 0.6061\times47.5/16$. The additional 
factor $47.5/16$ in $b_4$ is to account for the energy and pressure 
contributions from the additional quark degrees of freedom compared to pure 
$SU(3)$ case. With the above values, $l\left(x\right)\longrightarrow y = 
b_{3}/2+\frac{1}{2}\times \sqrt{b_{3}^{2} + 4b_{2}\left(T=\infty\right)}$ as 
$T \longrightarrow \infty$. $l\left(x\right)$ and other quantities are then 
normalized as follows,
\begin{equation}
l\left(x\right) \rightarrow \frac{l\left(x\right)}{y}, ~~ b_{2} \rightarrow 
\frac{b_{2}}{y^{2}}, ~~ b_{3} \rightarrow \frac{b_{3}}{y}, ~~ b_{4} \rightarrow 
b_{4}y^{4},
\end{equation}
so that $l\left(x\right) \longrightarrow 1$ as $T \longrightarrow \infty$. The 
normalized quantities are then used in eqn. (\ref{eq:pispot}), which is then 
used to calculate the $l\left(x\right)$ profile using energy minimization, see 
ref.\cite{Layek:2005fn} for details. Fig. (\ref{fig:lprfl}) shows the plot of 
$|l(x)|$ for the interface between two different vacua at T = 400 MeV (in 
the absence of quarks all the three interfaces have same profile for $|l(x)|$). 

An interpolating $l(x)$ profile between different $Z(3)$ vacua, essentially 
implies that there is a background gauge field $A_{0}(x)$ profile which 
interpolates between different $Z(3)$ vacua. (This is an important assumption
for our work, and also for refs.
\cite{KorthalsAltes:1992us,KorthalsAltes:1994be}). As the order parameter is
the thermal expectation value of the Polyakov loop, its relation to any
underlying gauge field configuration is not direct. The assumption of
a time independent background $A_0$ field directly determined via
eqn.(\ref{eq:a0diag}) is a simple choice, and we take it in that spirit.)  
This spatial variation of $A_{0}$ gives rise to a localized color electric 
field in the QGP medium. The quarks/anti-quarks moving across the $Z(3)$ 
domain walls will behave differently in the presence of such (color) electric 
field configuration. As a result, we 
should have different reflection and transmission coefficient for quarks and
anti-quarks. This is the source of CP violation. (This CP  asymmetry is
spontaneous because it arises from a specific configuration of the 
background $A_{0}$ field, which manifests itself as a potential in the 
equation of motion for quarks/antiquarks.) The earlier studies 
\cite{KorthalsAltes:1992us,KorthalsAltes:1994be} of this spontaneous
CP violation arising from Z(3) walls focused on the localized solution
of Dirac equation (in Euclidean space), and it was shown that if a
wave function for a fermion species localizes, then it's CP conjugate
doesn't. It was also showed in ref.\cite{KorthalsAltes:1994if} that in the 
Standard Model and Minimally Supersymmetric Extension of the Standard 
Model, this CP violation can be utilized via sphaleron processes to lead 
to baryogenesis in the early universe. 

The background gauge potential $A_0$ associated with the profile
of $l(x)$  was first calculated by us in ref. \cite{Atreya:2011wn} where 
we also discussed various conceptual issues related to the ambiguities in
the extraction of a colored quantity $A_0$ from color singlet $l(x)$. We 
choose Polyakov gauge (diagonal gauge) for $A_{0}$ :

\begin{equation} 
A_{0} = \frac{2\pi T}{g}\left(a\lambda_{3} + b\lambda_{8}\right),
\label{eq:a0diag}
\end{equation}

where, $g$ is the coupling constant and $T$ is the temperature, while 
$\lambda_{3}$ and $\lambda_{8}$ are the diagonal Gell-Mann matrices.
The $A_{0}$ profile was obtained from $l(x)$ profile 
(Fig. \ref{fig:lprfl}) by inverting eqn.(\ref{eq:lx}). We also calculated 
reflection and transmission coefficient of quarks and anti-quarks and it was 
found that the CP violating effect was stronger for heavier quarks. 
For details, see refs. \cite{Atreya:2011wn}.

\section{FORMATION OF $Z(3)$ DOMAIN WALLS IN THE EARLY UNIVERSE}
\label{sec:z3form}
   
 The possible mechanism for the formation of these $Z(3)$ domain walls in 
the early universe was discussed in detail in \cite{Layek:2005zu}. We briefly
recall essential points from that discussion. One important difference
for the formation of Z(3) walls compared to the formation of
other topological defects in the early universe arises from the fact
that here symmetry is broken in the high temperature phase, and is restored
as the universe cools while expanding. Standard mechanism of formation
of defects (the Kibble mechanism) leads to the formation of defects during 
the transition to the symmetry broken phase. What happens when the universe
was already in the symmetry broken phase from the beginning? One could use
general arguments of causality etc. to get some bounds on Z(3) domain walls
but it is not satisfactory, especially in view of quark mass effect due
to which all domain walls can disappear (in principle, in a short time). 
To discuss the detailed formation of $Z(3)$ structures using standard defect
formation scenario, one would require a situation where the universe 
undergoes the transition from the hadronic (confined/low temperature) phase to 
the QGP (deconfined/high temperature) phase. Kibble mechanism 
\cite{Kibble:1976sj} can then  be invoked to study the formation of these 
defects.  As discussed in ref.\cite{Layek:2005zu}, inflationary cosmology 
provides a natural resolution of this problem as we discuss below. 
  
Before inflation, the universe was at a very high temperature ($T>>T_{c}$) and 
quarks and gluons were deconfined. During inflation, the temperature of the 
universe decreases exponentially to zero due to the rapid expansion. As a 
result any previously existing $Z(3)$ interfaces disappear as the temperature 
drops below the critical temperature $T_{c}$ (if universe remains in 
quasi-equilibrium during this period) or as the 
energy density drops below $\Lambda_{QCD}$ due to expansion (in a standard out 
of equilibrium scenario). After inflation, the 
universe starts reheating and eventually the temperature is higher than 
critical temperature for confinement-deconfined transition. During
the stage when temperature of the universe rises above the quark-hadron
transition, Z(3) symmetry will break spontaneously, and Z(3) walls and
associated QGP string will form via the standard Kibble mechanism.
For $T >> \Lambda_{QCD}$, the energy scale for 
these walls is set by the temperature of the universe. The tension of the 
$Z(3)$ interface and associated string \cite{Layek:2005fn} is set by the QCD 
parameters and the temperature. As a result the dynamics, of the tension 
forces at the least, should be decided by the background plasma for 
temperatures far above the QCD scale. However, in presence of quarks, there is 
an explicit breaking of $Z(3)$ symmetry. Two of the vacua, with $l(x)=z,~
z^{2}$, become metastable leading to a pressure difference between the true 
vacuum and the metastable vacua \cite{Dixit:1991et,KorthalsAltes:1994be}. This 
leads to a preferential shrinking of metastable vacua. As the collapse of 
these regions can be very fast (simulations indicate $v_{w} \sim 1$ 
\cite{Gupta:2010pp,Gupta:2011ag}), they are unlikely to survive until late 
times, say until QCD scale, to play any significant role in the context of the 
universe. However, there is a possibility that when effects of quarks 
scattering from the walls is taken into account their collapse may be slower 
due to the friction experienced by domain wall. For large friction, the walls 
may even remain almost frozen in the plasma. For example, it has been 
discussed in the literature that dynamics of light cosmic strings can be 
dominated by friction which strongly affects the coarsening of string network 
\cite{Chudnovsky:1988cv,Martins:1995tg}. It is  plausible that the dynamics of 
these $Z(3)$ 
walls is friction dominated because of the non-trivial scattering of quarks 
across the wall. This can lead to significant friction in wall motion.

  Even if the dynamics of the domain walls is not strongly friction 
dominated, it is still possible for these $Z(3)$ domains to survive until the 
QCD scale, in 
certain low energy inflationary models 
\cite{Knox:1992iy,Copeland:2001qw,vanTent:2004rc}. In these models the 
reheating temperature can be quite low ($\sim 1~TeV$, or preferably,
even lower)). With inclusion of some friction in the dynamics of domain 
walls, it is then possible for the walls to survive until QCD transition. 
Note that the pressure difference between the true vacuum and metastable 
vacuum may affect the formation of these domains. For example, there may be 
a bias in formation of these domains as temperature crosses $T_{c}$ due to 
this pressure difference. Though such a bias may get  washed out 
by the thermal fluctuations and the continued rapid reheating at the
end of inflation when equilibrium concepts may not strictly apply. It is
also possible that the pressure difference between the metastable $Z(3)$ 
vacua and the true vacuum resulting from the explicit symmetry breaking 
term may be small near $T_c$. We will assume such optimistic conditions
to apply and  continue to use the effective potential given in 
eqn. (\ref{eq:pispot}) for the rest of the discussion, ignoring the 
effects of explicit symmetry breaking due to quarks. For detailed
discussion of these issues regarding formation of Z(3) walls in
the early Universe see ref.\cite{Layek:2005zu}. Certainly it is important to
consider the validity of these assumptions in detail, e.g. the evolution
of domain walls with due account of friction due to quark-gluon scatterings,
and we hope to come back to this in future.

After formation, the domain wall network undergoes coarsening, leading to  
only a few domain walls within the horizon volume. 
Basically with our assumptions
of neglect of explicit symmetry breaking due to quarks, the standard
scaling distribution of domain walls will be expected, with few domain
walls surviving within horizon at any stage. Detailed simulation of the 
formation and evolution of these $Z(3)$ walls in the context of RHICE is 
discussed in ref \cite{Gupta:2010pp,Gupta:2011ag}. Even though 
the simulations are done with first order transition via bubble nucleation,
resulting domain wall network is reasonably independent of that. This is
because the basic physics of the Kibble mechanism only requires formation of
uncorrelated domains which happens in any transition. Further, the evolution 
of these $Z(3)$ domain walls, once they are formed, can be understood quite 
well from these simulations. As we discussed previously, large 
friction due to quark scatterings can lead to slow dynamics of walls (with 
negligible wall velocities) and may help in retaining large sizes upto the 
stage of quark-hadron transition. (Simulation in ref.
\cite{Gupta:2010pp,Gupta:2011ag} did not
take into account of the friction due to scattering by quarks and gluons,
though dissipation due to fluctuations of the Polyakov loop order parameter
was automatically included.)  

\section{GENERATION AND EVOLUTION OF BARYON INHOMOGENEITIES}
\label{sec:binhm}
     
   In this section we discuss how these collapsing $Z(3)$ walls lead to 
the segregation of baryon number leading to the formation of quark 
and antiquark nuggets. After the domain walls have formed (as discussed 
in the previous section), the closed domains start to collapse. (Again,
with neglect of explicit symmetry breaking effects, otherwise even a closed
domain wall may expand depending on the pressure difference on the two
sides of the wall.)  As discussed in section \ref{sec:z3}, a non-trivial 
profile of $l(x)$ leads to  a background $A_{0}$ profile. This $A_{0}$ will 
interact with quarks and anti-quarks in a different manner. In other words, 
it will have different reflection and transmission coefficients for the 
quarks and antiquarks leading to a spontaneous violation of CP symmetry. 
This will lead to the concentration of quarks (or anti-quarks, depending on
the wall) within the collapsing domain, 
thereby resulting in the segregation of baryons and anti-baryons in the early 
universe. These collapsing baryon (anti-baryon) rich regions can form quark 
(anti-quark) nuggets if the baryon concentration is sufficiently high in these 
regions. It is important to note that these $Z(3)$ walls exist in the QGP
phase as topological defects, forming irrespective of the order of the 
quark-hadron phase transition, even if it is a cross-over. Hence, the 
formation of quark nuggets in our model is via 
a very different mechanism than the originally proposed 
one \cite{Witten:1984rs}. In context of $Z(3)$ walls, the baryon inhomogeneity 
generation was discussed by some of us in ref. \cite{Layek:2005zu}. However 
there was no CP violation in that discussion as it dealt with only $l(x)$ 
profile and not the gauge field associated with $l(x)$.

  Main aspects of calculations in ref.\cite{Layek:2005zu} were along the line
of ref.\cite{Fuller:1987ue}. We continue to follow that approach here.
While studying the baryon transport across the domain wall, we assume 
constant temperature. A major simplification that happens due to this 
assumption is that one can take the height of the potential to be constant. 
This also makes it possible for us to ignore the effects coming from the 
reheating due to decreasing surface area as the wall collapses. We also 
assume that the thermal equilibrium is maintained as the quarks and 
antiquarks are reflected from the domain wall. We further assume that the 
collapse of the domain walls is fast. This allows us to ignore the 
expansion of the universe as domain walls will then collapse in the time 
smaller than the Hubble time. In our calculations we take the wall velocity to 
be the sound velocity, $v_{w}=1/\sqrt{3}$. This velocity could be larger if 
the friction is subdominant in comparison to the surface tension of the wall,
or the velocity can be much smaller if the frictional forces are very dominant. 
To study the change in the number densities inside and outside the collapsing 
region we assume that the baryons homogenize instantaneously as the baryon 
transport occurs across the wall (See the discussion in 
ref.\cite{Fuller:1987ue} for the self consistency of this assumption). We can 
then work with only the number density inside and outside the domain wall and 
ignore the diffusion of baryons.

  Let $V$ be the Hubble volume at time $t$. In this volume suppose there are 
$N_{d}$ number of collapsing domains. Let $V_{i} = 4\pi/3R(t)^{3}N_{d}$ 
($R(t)$ being the radius of domain taken to be spherical) be the volume 
contained within the domain 
walls and $V_{o} = V-V_{i}$ be the volume outside the collapsing regions.
As we are ignoring the expansion of the universe for a given domain wall, 
$V$ is fixed. Note that this assumption here amounts to saying that for
a reasonably large value of $N_d$, and with large wall velocity, the collapse
of domain walls happens in a time much shorter than the Hubble time.

The radius of the collapsing domain, at time $t$, is given by the expression
\begin{equation}\label{eq:radius}
R(t) = \frac{r_{H}}{N_{d}^{1/3}} - v_{w}(t-t_{0}),
\end{equation}
where $r_{H}$ is the horizon size at the initial time $r_H \simeq 
t_{0}\simeq 30\left( 
\frac{150}{T(MeV)}\right)^{2}$ (in the units of micro seconds). If $n_{i}$ and 
$n_{o}$ are the number densities of baryons in the regions inside and outside 
the domain walls, then the total number of baryons in each region is $N_{i} = 
n_{i}V_{i}$ and $N_{o} = n_{o}V_{o}$. The equations for studying quark number 
density concentration inside and outside the domain wall can then be written 
as
\begin{equation}
\dot{n_{i}} = \Bigl(-\frac{2}{3}v_{w}T_{w}n_{i} + \frac{v_{o}^{rel}n_{o}T_{-}- 
v_{i}^{rel}n_{i}T_{+}}{6}\Bigr)\frac{S}{V_{i}} - n_{i}\frac{\dot{V_{i}}}{V_{i}} \\
\label{eq:numin}
\end{equation}
\begin{equation}
\dot{n_{0}} = \Bigl(\frac{2}{3}v_{w}T_{w}n_{i} - \frac{v_{o}^{rel}n_{o}T_{-}- 
v_{i}^{rel}n_{i}T_{+}}{6}\Bigr)\frac{S}{V_{i}} + n_{o}\frac{\dot{V_{i}}}{V_{o}}, \\
\label{eq:numout}
\end{equation}
where $S$ is the surface area of the collapsing wall. $T_{w}$ is the 
transmission coefficient for the quarks inside the domain and moving parallel 
to the wall. The relative velocity for such quarks with respect to the wall is 
$v_{w}$ and they constitute $4/6$ of the total number of the inside quarks. 
$T_{-} ~\left(T_{+}\right)$ is the transmission coefficient calculated for the 
quarks that are moving from outside (inside) of the wall towards the inside 
(outside) with the relative velocity $v_{o}^{rel}~\left(v_{i}^{rel}\right)$ 
with respect to the wall. Each contributes towards $1/6$ of the corresponding 
number densities. Eqn. (\ref{eq:radius}), (\ref{eq:numin}) and 
(\ref{eq:numout}) are then solved simultaneously to get the evolution of the 
baryon densities inside the collapsing domain.

  As the wall collapses, it leaves behind a profile of baryon density. Consider 
a spherical shell of thickness $dR$, at a distance $R$ from the center of the 
domain wall. Then if $\rho\left(R\right)$ is the baryon density, then total 
number of baryons in the shell is given by $dN_{i} = 4\pi R^{2}\rho\left(R 
\right)dR$. Using eqn (\ref{eq:radius}) we get,
\begin{equation}
 \rho\left(R\right) = -\frac{\dot{N_{i}}}{4\pi v_{w}R^{2}}.
\label{eq:rho}
\end{equation}
Eqn. (\ref{eq:radius}) and (\ref{eq:rho}) are solved simultaneously to get the 
density profile. It is important to note that during last stages of the 
collapse of domain wall, it is possible that the baryon concentration becomes 
so large that chemical potential in the region is comparable to the 
temperature. This will alter the transmission probability of the baryons across 
the domain wall. We are neglecting any such effects that may arise during the 
evolution. 

   As we discussed in section \ref{sec:z3}, the domain wall is selective in 
the transmission of baryons and anti-baryons due to its CP odd nature. This 
will lead to the baryon anti-baryon segregation. As a result we get baryon 
rich and anti-baryon rich regions that can form nuggets and anti-nuggets if 
there is sufficient concentration of baryons or anti baryons. In addition, the 
domain wall is also sensitive to the color of quark as it has different 
reflection and transmission coefficient for different colors. Eqn 
(\ref{eq:radius}) to (\ref{eq:rho}) need to be solved for each color which 
will result in the color specific baryon concentration. This in itself is not 
surprising as in the QGP phase, the degrees of freedom are color degree of 
freedom and the requirement to have colorless objects in QGP would be an 
artificial one. 

\section{RESULTS}
\label{sec:results}

\begin{figure}[!htp]
\begin{center}
\includegraphics[width=0.45\textwidth]{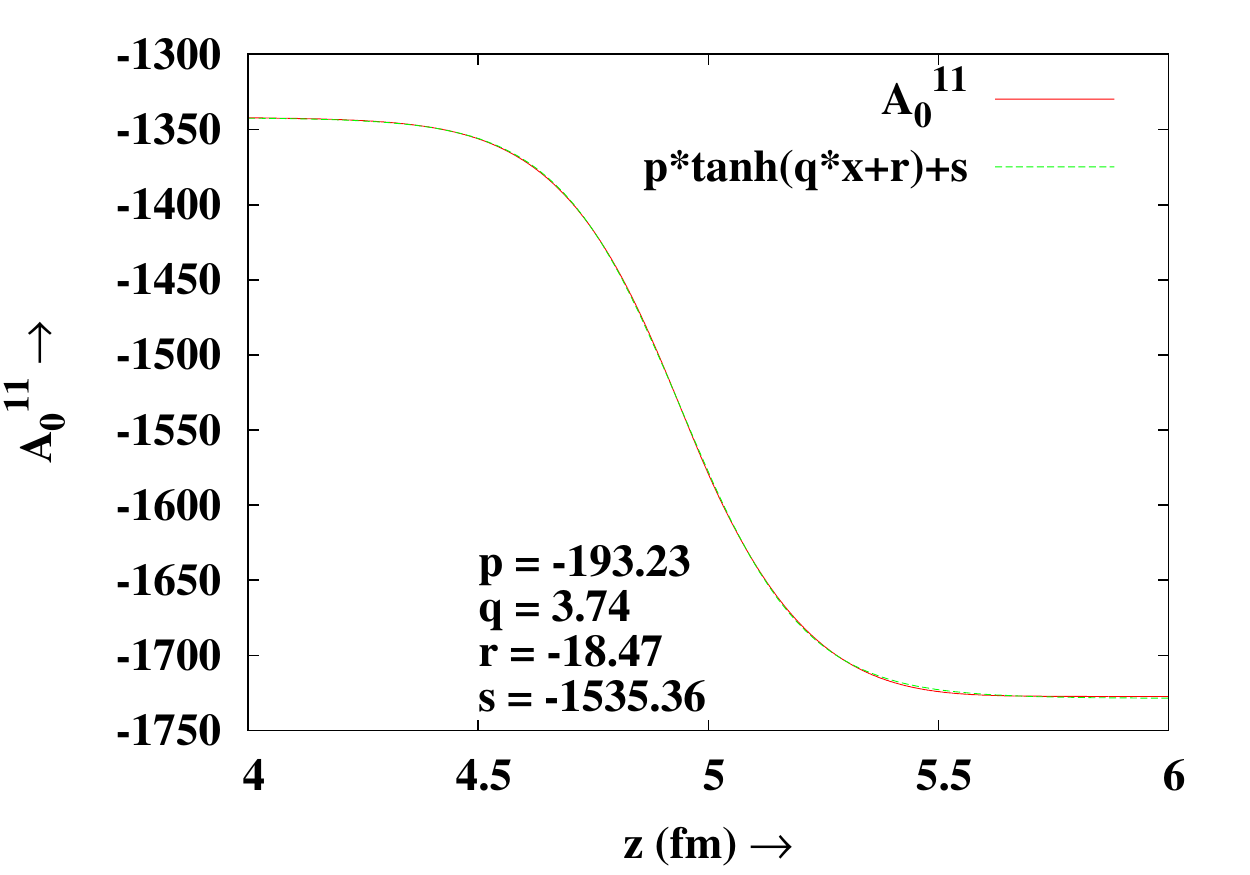}
\includegraphics[width=0.45\textwidth]{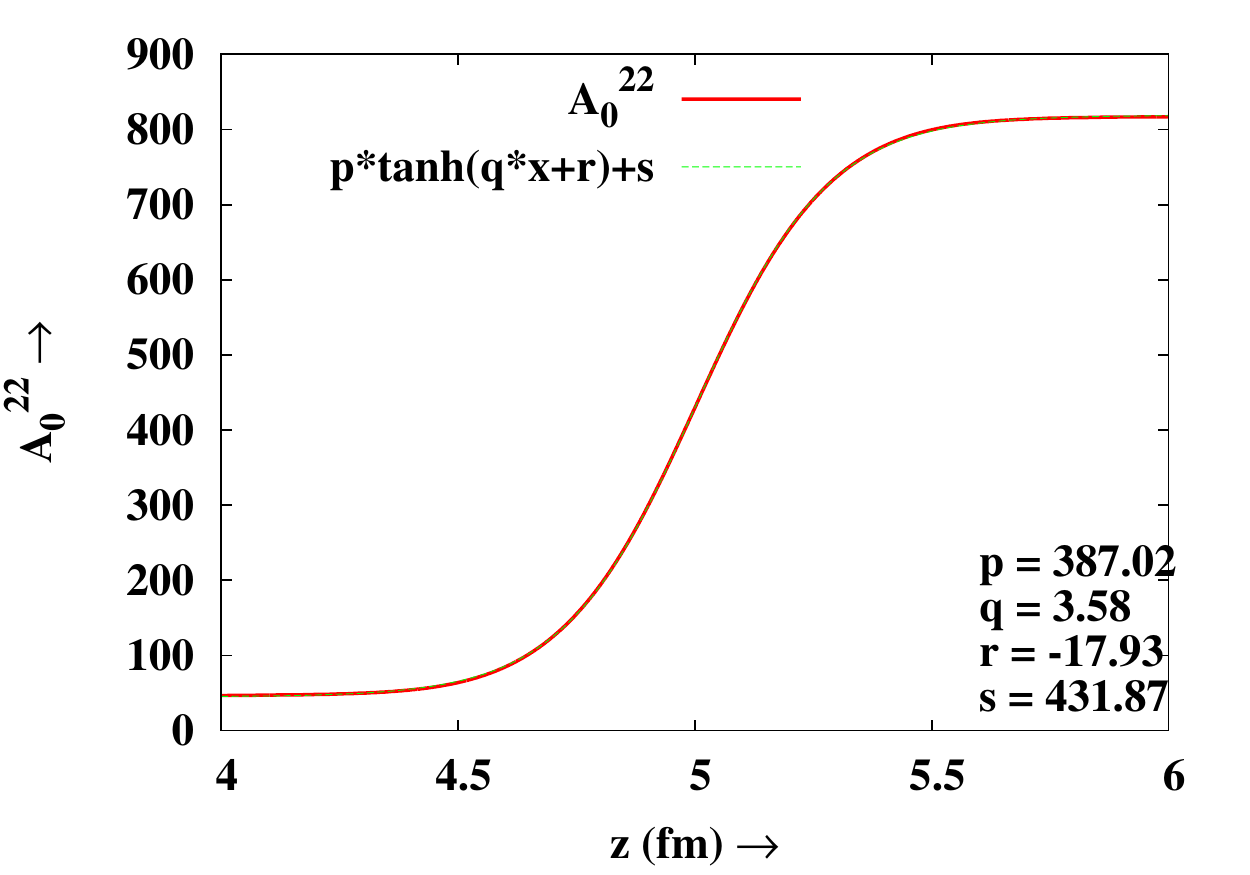}
\includegraphics[width=0.45\textwidth]{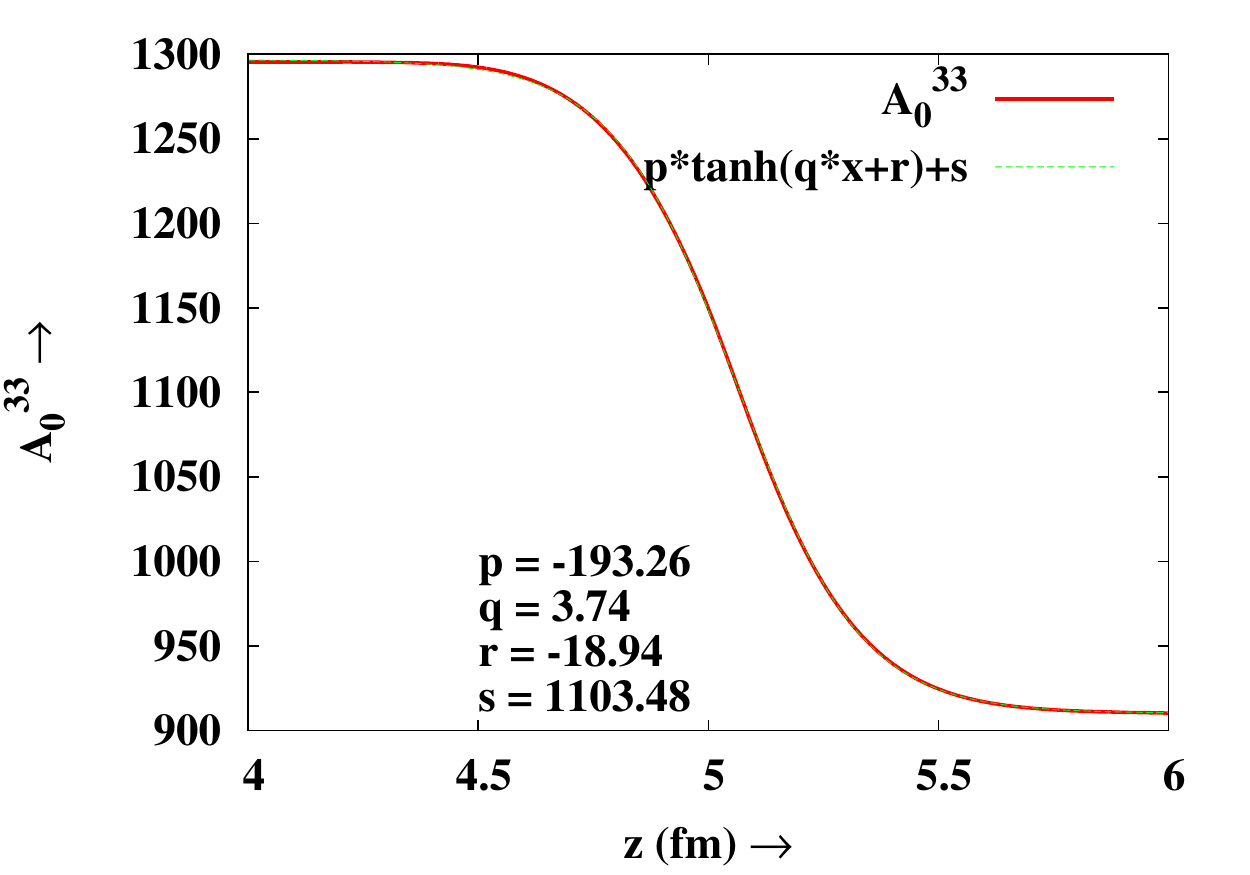}
\caption{The background $A_{0}$ profile calculated from the $l(x)$ profile. The
 profile is fitted to a $\tanh$ curve.}
\label{fig:a0plt}
\end{center}
\end{figure}

  We present a brief discussion of  how to obtain $A_{0}$ profile from 
$l(x)$ profile. See ref. \cite{Atreya:2011wn} for details. Substituting 
eqn.(\ref{eq:a0diag}) in eqn. (\ref{eq:lx}) and comparing the 
real and imaginary parts, we get
\begin{subequations}
\begin{align}
\cos\left(\alpha\right) + \cos \left(\beta\right)
    + \cos \left(\gamma\right)& = 3 |l(x)| \cos \left(\theta \right), \\
\sin\left(\alpha\right) + \sin \left(\beta\right)
    + \sin \left(\gamma\right)& = 3 |l(x)| \sin \left(\theta \right),
\end{align}
\label{eq:lgrp}
\end{subequations}

where, $\alpha = 2\pi\left(\frac{a}{3} + \frac{b}{2}\right)$ ,
$\beta = 2\pi\left(\frac{a}{3} - \frac{b}{2}\right)$ and $\gamma
= 2\pi(\frac{-2a}{3})$ ((a,b) are defined in Eq.(7)).
$\theta$ is defined by $l(x) = |l(x)| e^{i\theta}$. For each of the 
$l = 1,z,z^{2}$ vacuum, the solutions are a set of ordered pairs 
$\left(a,b\right)_{L=1,z,z^{2}}$. We choose one pair $\left(a,b\right)_{L=1}$ as 
the initial condition. By demanding that $a$ and $b$ (and hence $A_{0}$) 
vary smoothly across the wall (as the profile of $L(x)$ changes smoothly), we 
approach the appropriate values of $\left(a,b\right)_{L=z}$ in $L = z,z^{2}$ 
vacuum. Once, we have $a$ and $b$ profiles, $A_{0}$ was calculated using eqn. 
(\ref{eq:a0diag}). Fig (\ref{fig:a0plt}) shows the background $A_{0}$ profile 
between $l=1$ and $l=z^{2}$, calculated using the profile given in fig. 
(\ref{fig:lprfl}) for $T = 400$ MeV.

  To calculate the reflection and transmission coefficient, we need the 
solutions of Dirac equation in the Minkowski space but the $A_{0}$ profile is 
calculated in Euclidean space. We start with the Dirac equation in the 
Euclidean space, with the spatial dependence of $A_0$ calculated from Z(3) 
wall profile as mentioned above. Then we do the analytic continuation of the 
full equation to the Minkowski space and use the resulting equation to 
calculate the reflection and transmission coefficients. We first approximated 
domain wall by the step potential. For a general smooth potential we 
followed a numerical approach given by Kalotas and Lee \cite{Kalotas:1991kl}. 
They have discussed a numerical technique to solve Schr\"{o}dinger equation 
with potentials having arbitrary smooth space dependence. We applied this
technique of ref.\cite{Kalotas:1991kl} for solving the Dirac equation. 

\begin{figure}[!htp]
\begin{center}
\includegraphics[width=0.5\textwidth]{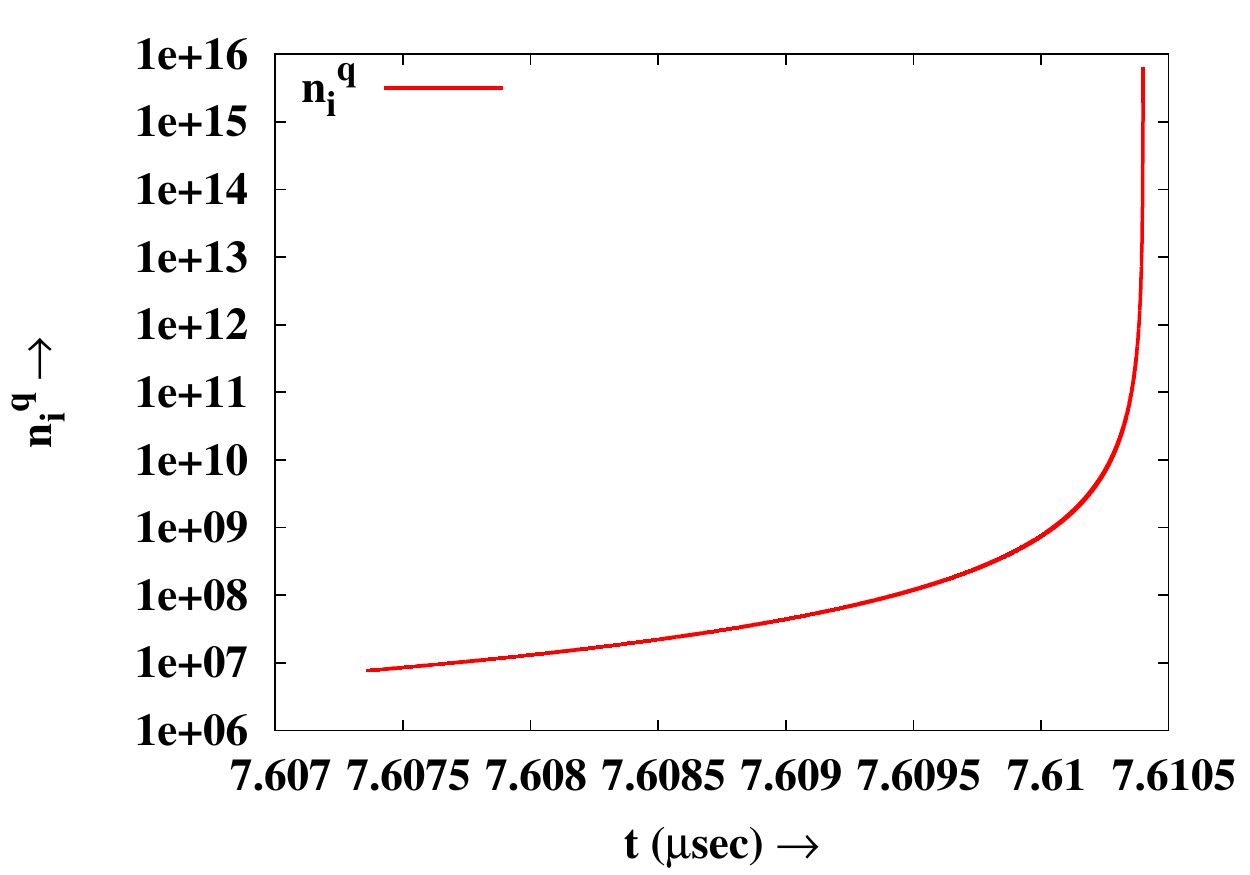}
\includegraphics[width=0.5\textwidth]{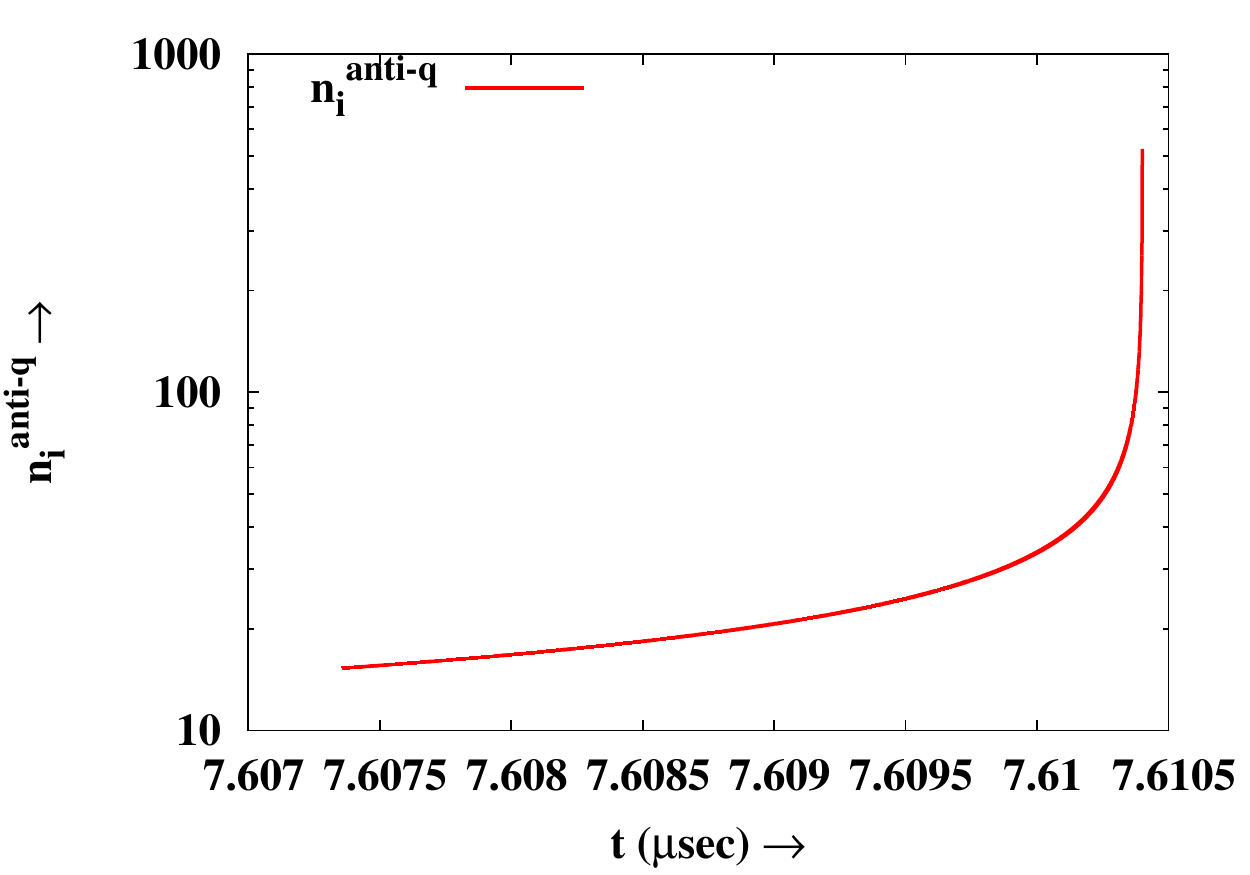}
\caption{Number density evolution with step function profile: (a)For Red, 
green and anti-blue charm quark. (b)For anti-red, anti-green and blue charm 
quark.}
\label{fig:numstp}
\end{center}
\end{figure}

\begin{figure}[!htp]
\begin{center}
\includegraphics[width=0.5\textwidth]{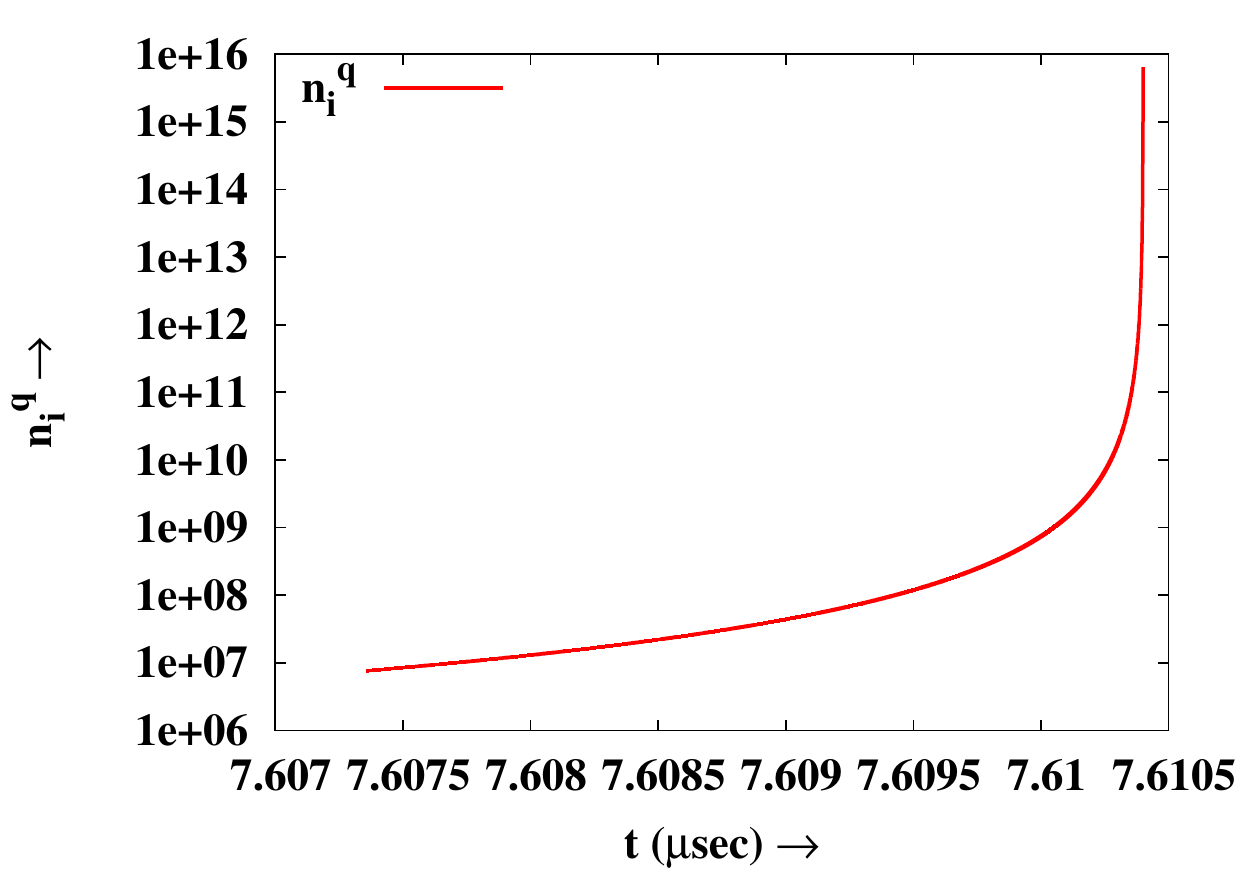}
\includegraphics[width=0.5\textwidth]{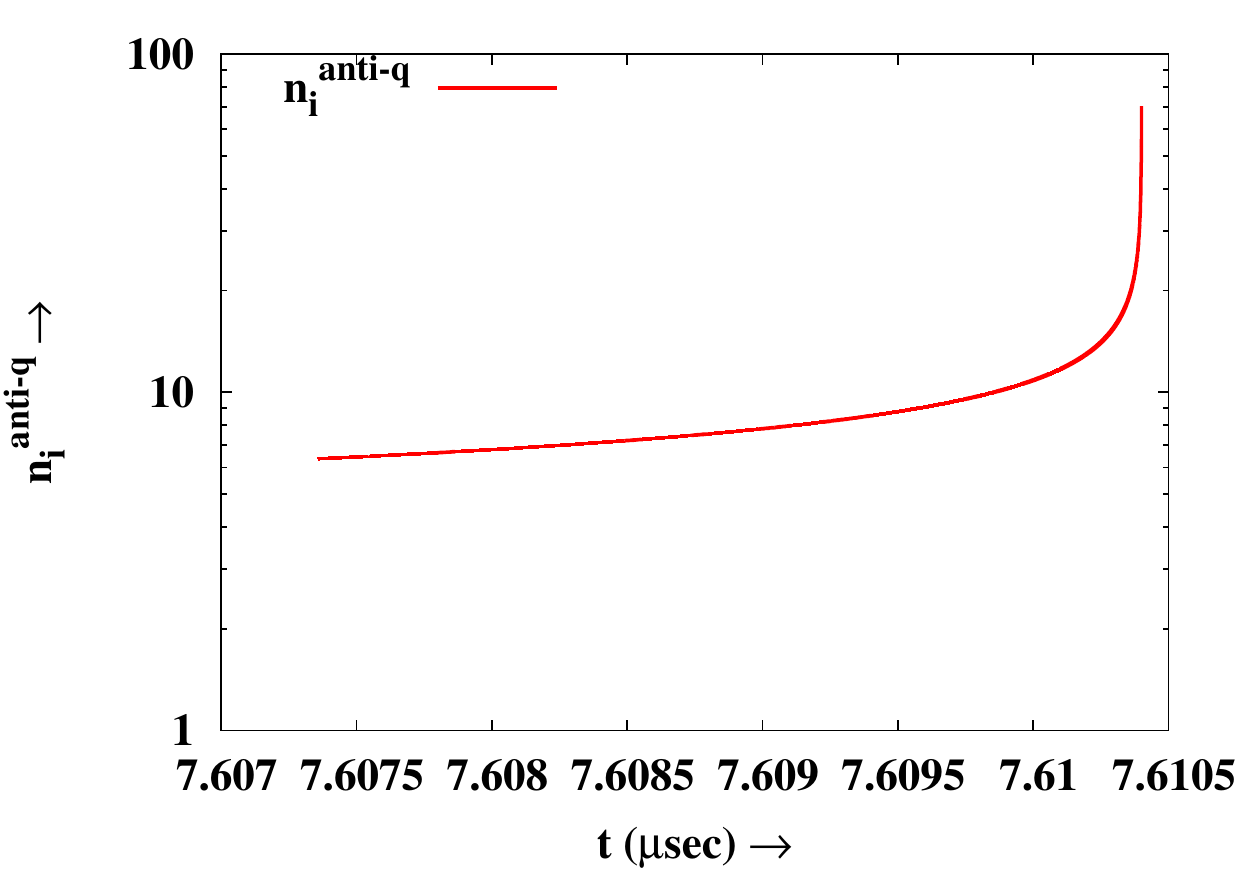}
\caption{Number density evolution with smooth profile: (a)For Red, 
green and anti-blue charm quark. (b)For anti-red, anti-green and blue charm 
quark.}
\label{fig:numsmth}
\end{center}
\end{figure}

\begin{figure}[!htp]
\begin{center}
\begin{tabular}{ccc}
\includegraphics[width=0.5\textwidth]{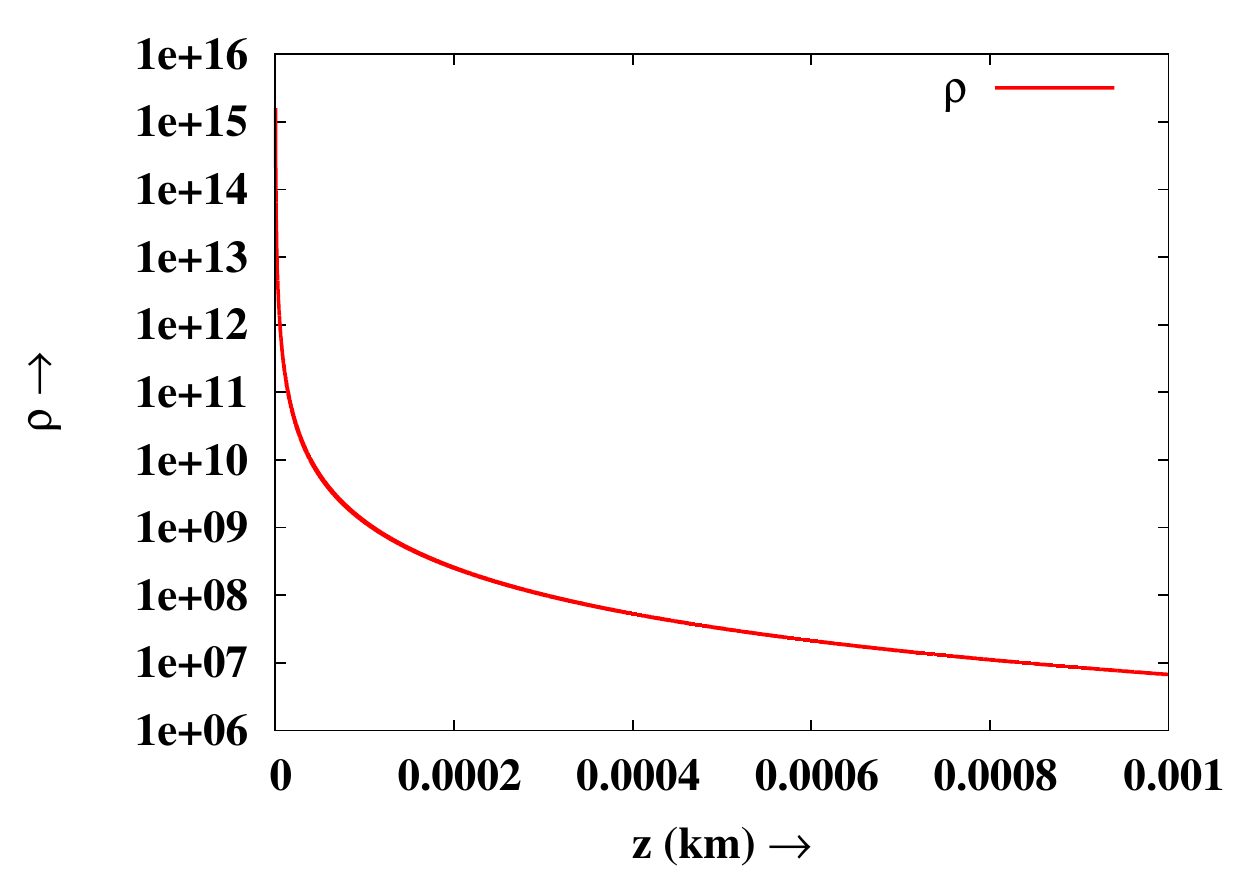}
\end{tabular}
\caption{Evolution of baryon density profile}
\label{fig:rho}
\end{center}
\end{figure}

 We will discuss the concentration of charm quarks in the following. Their
number density at $T \simeq 400$ MeV is still significant and with large
reflection coefficients, they lead to large baryon/anti baryon concentrations.
Up and down quarks are ultra-relativistic and have very small reflection
coefficients. The case of strange quark is an important one. We will comment 
on that case at the end of this section.  For charm quark at $T=400~MeV$, 
the thermal velocity $v_{p}$ is less than 
the sound velocity $v_{s}$. As we are assuming the wall velocity $v_{w}$ to be
same as $v_{s}$, the particles moving from outside towards the wall are unable 
to catch up. This means that $T_{+}$ is identically zero. For the particles 
moving towards the wall, the energy (in the rest frame of the wall) is much 
larger than the potential so most of them pass through ($T_{-}$ is close to 
unity). Only the particles moving parallel to the wall can get concentrated. 
The potential as seen by the incoming fermion is $V(z) = -gA_{0}(z)$. The 
value of $g$ is chosen such that $N/g^{2}=0.8$. Since $g$ is positive for 
quarks, the background $A_{0}$ profile dictates that red, green and anti-blue 
quarks are concentrated in the collapsing regions with $l=z^{2}$. (Note,
in Fig. \ref{fig:a0plt}, $A_0^{22}$ has opposite sign compared to $A_0^{11}$ 
and $A_0^{33}$.
Thus, while red and green quarks experience a potential barrier leading to
significant reflection, the blue quark sees a potential well. It is the
blue anti-quark which experiences a potential barrier and undergoes 
significant reflection.) Table  (\ref{tab:trns}) lists the values of 
$T_{w}$ for charm quark for smooth profile. It clearly indicates that 
two color species of quarks and one color species of 
anti-quark are not transmitted.
\begin{table}[!htp]
\begin{center}
\begin{tabular}{|c|c|c|c|}
\hline
 & $r$ & $b$ & $g$ \\
\hline
$c\quad$ & $0.0$ & $0.936623$ & $0.0$\\
\hline
$\bar{c}\quad$ & $0.997471$ & $0.0$ & $0.99903$\\
\hline
\hline
\end{tabular}
\caption{Table for the transmission coefficients for charm quarks and 
anti-quarks, moving parallel to the wall, from the $l=z^{2}$ wall.}
\label{tab:trns}
\end{center}
\end{table}
  These transmission coefficients were then used to solve eqn. \ref{eq:numin} 
and \ref{eq:numout} simultaneously. This gives us the evolution of number 
densities inside and outside the domain wall for each color. 
Fig. \ref{fig:numstp}(a) and \ref{fig:numstp}(b) show the evolution of number 
densities for charm quark and 
anti-quark inside the collapsing domain wall at $T = 400 MeV$ for the case of 
step potential approximation. The result is for $N_{d} = 10$. It is clear that 
the number of quarks contained in the domain wall is several orders of 
magnitude higher than the number of anti-quarks. The number densities of 
quarks and anti-quarks are shown in fig \ref{fig:numsmth}(a) and 
\ref{fig:numsmth}(b). Looking at fig. (\ref{fig:numstp}a) and fig 
(\ref{fig:numsmth}a) we note that 
the number densities are not much different for the smooth and step potential. 
This might seem surprising. However a look at fig. (\ref{fig:numstp}b) and 
fig (\ref{fig:numsmth}b) clearly shows that the number density of anti-red 
(and other corresponding) quarks, that are not getting concentrated, is much 
less for the smooth profile than the step potential. So, the number densities 
in fig (\ref{fig:numstp}a) and (\ref{fig:numsmth}a) have same order of 
magnitude but not same numbers.
  Fig.\ref{fig:rho} shows the density profile of red charm quark. As the 
majority of anti-quarks are completely transmitted, they do not leave any 
density profile behind.

 Fig.(\ref{fig:numstp}) and (\ref{fig:numsmth}) give the number density of 
quarks in units 
of the background quark/antiquark number density $n_0$, as a function of 
the size of the collapsing domain wall. At T = 400 MeV, 
$n_0 \simeq O(1)/fm^3$ for each type of quark. This gives the net baryon 
number trapped inside the domain wall to be of order $10^{52}$ when 
domain wall collapses to a size of order one
meter. This is with the optimistic assumption that all the baryons
get trapped inside the wall while antiquarks leave the wall virtually
unreflected. This may not be a reasonable assumption, especially
in view of the assumption of thermal equilibrium and homogeneous
baryon distribution inside the wall. In the most conservative
scenario, the net baryon number inside the domain wall should remain
trapped. Net baryon number to entropy ratio being of order $10^{-10}$,
it is safe to say that at least net baryon number of order $10^{42}$
can be trapped inside collapsing domain walls. These {\it quark
nuggets} may then survive until present and provide dark matter. In this case 
(fig. \ref{fig:numsmth}a), we had a concentration of 
baryons. This concentration is due to the wall between $l(x) =1$ and 
$l(x) =z^{2}$ vacua. There would also be a wall between $l(x) =1$ and $l(x) = 
z$ vacua, which will be the conjugate of the wall between $l(x) =1$ and 
$l(x) =z^2$. In this domain, it will be the anti-baryons which will get 
concentrated. As a result we will have a net segregation of baryons and anti 
baryons. Though, note that for the concentration of antiquarks, the above
type of conservative estimate of $10^{42}$ baryon number may not be
applicable.

   An important point is the choice of initial conditions for calculating 
$A_{0}$. We will now discuss the effect of this choice of initial conditions on 
the baryon segregation. The ambiguity in the initial condition and hence in 
determining $A_{0}$ is reasonable as we are extracting information about a 
colored object ($A_0$) starting from a colorless variable $L(x)$. Thus there 
is no reason to expect unique solution for $A_0$ starting from a given $L(x)$ 
profile. This is reflected in the various sets $(a,b)$ that are available for 
each of the $Z(3)$ vacua. It appears  that choosing a different sets $(a,b)$ 
amounts to selecting domain wall profiles which carries different color 
information for the scattering of a fixed color (say red) quark (see 
ref. \cite{Atreya:2011wn} for a detailed discussion). In the 
present context that would simply mean that if for a specific choice 
of $(a,b)$, on color (say red) is being concentrated inside the collapsing 
domain, another color (say blue) will be concentrated in the region for a 
different choice of $(a,b)$. Nonetheless there would be concentration of 
quarks (or anti-quarks, as the case may be) and the number densities will also 
be same. 

   In his original proposal, Witten \cite{Witten:1984rs} discussed the 
formation of strangelets. We have not discussed concentration of strange
quarks. This is due to the fact that strange quarks are in 
Klein regime at these temperature i.e. reflection coefficients are greater 
than unity. As Klein paradox is understood in terms of particle anti-particle 
pair production, it seems likely that we will have even larger 
concentration of strange quarks (or anti-strange-quarks) because the pair 
produced species will 
also contribute to the number density inside the collapsing volume. However, 
there is a conceptual complication in doing the quantitative estimation of 
the number densities. In pair production, there is a back-reaction on the 
background field. The pair production is at the cost of the energy of the 
background field, which decreases as more and more particle are pair produced. 
This is difficult to implement in the present case as the background 
configuration is a topological configuration and it is not clear how to
decrease the magnitude of $A_0$ here (affecting the magnitude of $l(x)$)
while maintaining the topological property of the wall configuration.
Nonetheless, it is clear that the concentration of strange 
quarks/antiquarks of at least same order as above  will be expected in our
model, naturally leading to the formation of strangelets. This is one
of the strengths of our model that it can naturally lead to formation
of strangeness rich quark nuggets. As we mentioned in the introduction,
stability of strangelets has been discussed extensively in the literature
and for a wide range of quark numbers the strangelets could be stable.
From our discussion of the formation of Z(3) walls it is clear the formation
of small Z(3) walls is almost unavoidable in the QGP phase. Thus formation
of small strangelets will happen very naturally in our model. As we have
discussed above, under certain optimistic conditions, even very large
strangelets are possible within our model. 

\section{Discussions and Conclusions}
 \label{sec:disc}

 We have addressed the issue of viability of quark nuggets as dark matter
candidates by showing an alternate mechanism for the formation of
these objects in the QGP phase of the early universe. Here the nature,
or even the existence of quark-hadron phase transition is completely 
irrelevant. Quarks and antiquarks are reflected by collapsing Z(3) walls.
This leads to concentration of baryon number in localized regions, forming
quark nuggets, exactly as in the original scenario of Witten. This possibility
was discussed by some of us in an earlier paper \cite{Layek:2005zu} where
an effective constituent quark mass was introduced as a function of
the Polyakov loop order parameter. Here we have extended that analysis by
recognizing that the $A_0$ field associated with $l(x)$ leads to
spontaneous CP violation leading to different scattering of quarks and
antiquarks from a given Z(3) wall. Thus one gets quark nuggets as well
as antiquark nuggets in this scenario. Such nuggets and anti nuggets have 
been discussed in recent publications \cite{Lawson:2012zu} in context of a 
soft radio 
background. It would be interesting to explore if these nuggets and anti-
nuggets discussed here can play a role in such phenomenon. Importantly,
these nuggets and antinuggets provide a natural candidate, entirely
within the standard model, for dark matter of the universe.
    Note that as the CP violation here is resulting from a specific
domain wall configuration in a given region, overall there will not
be any net concentration of baryons or antibaryons. It is tempting to
speculate that with the use of CP violating $\theta$ term in the QCD 
Lagrangian, can one get a net concentration of antibaryons over 
antibaryons?
If that could be achieved then one can attempt to explain baryogenesis
also in this model where excess antibaryons remain trapped in antiquark
nuggets while compensating baryon number accounts for the visible
matter in the universe.

\section*{Acknowledgments}

 We thank Sanatan Digal, Saumia P. S., Partha Bagchi and Arpan Das for very 
valuable comments and useful discussions.

\section*{References}


\end{document}